\documentclass[
floatfix,
 reprint,
 amsmath,amssymb,
 aps,
]{revtex4-1}
\usepackage{bm}
\usepackage{braket}
\usepackage[dvipdfmx]{graphicx}
\usepackage{color}

\newcommand{\qq}{\Braket{\bar{q}q}}
\newcommand{\pim}{\pi^-}
\newcommand{\helium}{{}^3{\rm He}}

\newcommand{\Sn}[1]{{}^{#1}{\rm Sn}}
\newcommand{\reaction}{(d,\helium)}

\begin{document}
\title{
  Chiral symmetry restoration at high matter density observed in pionic atoms
}
\author{Takahiro~Nishi$^{1}$}
\author{Kenta~Itahashi$^{1,2}$}
\email{Email: itahashi@a.riken.jp}

\author{DeukSoon~Ahn$^{1,3}$}
\author{Georg~P.A.~Berg$^{4}$}
\author{Masanori~Dozono$^{1}$}
\author{Daijiro~Etoh$^{5}$}
\author{Hiroyuki~Fujioka$^{6}$}
\author{Naoki~Fukuda$^{1}$}
\author{Nobuhisa~Fukunishi$^{1}$}
\author{Hans~Geissel$^{7}$}
\author{Emma~Haettner$^{7}$}
\author{Tadashi~Hashimoto$^{2,8}$}
\author{Ryugo~S.~Hayano$^{9}$}
\author{Satoru~Hirenzaki$^{10}$}
\author{Hiroshi~Horii$^{9}$}
\author{Natsumi~Ikeno$^{11}$}
\author{Naoto~Inabe$^{1}$}
\author{Masahiko~Iwasaki$^{1,2}$}
\author{Daisuke~Kameda$^{1}$}
\author{Keichi~Kisamori$^{12}$}
\author{Yu~Kiyokawa$^{12}$}
\author{Toshiyuki~Kubo$^{1}$}
\author{Kensuke~Kusaka$^{1}$}
\author{Masafumi~Matsushita$^{12}$}
\author{Shin'ichiro~Michimasa$^{12}$}
\author{Go~Mishima$^{9}$}
\author{Hiroyuki~Miya$^{1}$}
\author{Daichi~Murai$^{1}$}
\author{Hideko~Nagahiro$^{10}$}
\author{Megumi~Niikura$^{9}$}
\author{Naoko~Nose-Togawa$^{13}$}
\author{Shinsuke~Ota$^{12}$}
\author{Naruhiko~Sakamoto$^{1}$}
\author{Kimiko~Sekiguchi$^{5}$}
\author{Yuta~Shiokawa$^{5}$}
\author{Hiroshi~Suzuki$^{1}$}
\author{Ken~Suzuki$^{7,14}$}
\author{Motonobu~Takaki$^{12}$}
\author{Hiroyuki~Takeda$^{1}$}
\author{Yoshiki~K.~Tanaka$^{2}$}
\author{Tomohiro~Uesaka$^{1}$}
\author{Yasumori~Wada$^{5}$}
\author{Atomu~Watanabe$^{5}$}
\author{Yuni~N.~Watanabe$^{9}$}
\author{Helmut~Weick$^{7}$}
\author{Hiroki~Yamakami$^{6}$}
\author{Yoshiyuki~Yanagisawa$^{1}$}
\author{Koichi~Yoshida$^{1}$}
\affiliation{\it $^{1}$ RIKEN Nishina Center for Accelerator-Based Science, RIKEN, Saitama, Japan}
\affiliation{\it $^{2}$ RIKEN Cluster for Pioneering Research, RIKEN, Saitama, Japan}
\affiliation{\it $^{3}$ Center for Exotic Nuclear Studies, Institute for Basic Science (IBS), Daejeon, Republic of Korea}
\affiliation{\it $^{4}$ Department of Physics and the Joint Institute for Nuclear Astrophysics Center for the Evolution of the Elements, University of Notre Dame, Indiana, USA}
\affiliation{\it $^{5}$ Department of Physics, Tohoku University, Sendai, Japan}
\affiliation{\it $^{6}$ Department of Physics, Kyoto University, Kyoto, Japan}
\affiliation{\it $^{7}$ GSI Helmholtzzentrum f\"{u}r Schwerionenforschung GmbH, Darmstadt, Germany}
\affiliation{\it $^{8}$ Advanced Science Research Center, Japan Atomic Energy Agency, Ibaraki, Japan}
\affiliation{\it $^{9}$ Department of Physics, School of Science, The University of Tokyo, Tokyo, Japan}
\affiliation{\it $^{10}$ Department of Physics, Nara Women's University, Nara, Japan}
\affiliation{\it $^{11}$ Department of Life and Environmental Agricultural Sciences, Faculty of Agriculture, Tottori University, Tottori, Japan}
\affiliation{\it $^{12}$ Center for Nuclear Study, the University of Tokyo, Saitama, Japan}
\affiliation{\it $^{13}$ Research Center for Nuclear Physics, Osaka University, Osaka, Japan}
\affiliation{\it $^{14}$ Ruhr-Universit\"at Bochum, Bochum, Germany}
\collaboration{piAF Collaboration}

\date{\today}

\maketitle

{\bf According to quantum chromodynamics, the vacuum is not an empty space
as it is filled with quark-antiquark pairs. The pair
has the same quantum numbers as the vacuum and forms a condensate
because the strong interaction of the quantum chromodynamics is too
strong to leave the vacuum empty. This quark-antiquark condensation,
the chiral condensate,
breaks the chiral symmetry of the vacuum. The expectation value of the
chiral condensate is an order parameter of the chiral symmetry, which
is expected to decrease at high temperatures or high matter densities
where the chiral symmetry is partially restored. Head-on collisions
of nuclei at ultra-relativistic energies have explored
the high temperature regime but experiments at high
densities are rare. Here, we measure the spectrum of pionic \boldmath$\Sn{121}$
atoms and study the interaction between the pion and the nucleus. We
find that the expectation value of the chiral condensate is reduced at
finite density compared to the value in vacuum. The reduction is
linearly extrapolated to the nuclear saturation density and indicates
that the chiral symmetry is partially restored due to the extremely
high density of the nucleus.}

\section*{Chiral Symmetry and Pion-Nucleus Interaction}
The properties of the vacuum depend on
the temperature and the matter density~\cite{Weise93,Brown96}.
In the present universe, the low energy density makes the vacuum lose the
chiral symmetry, and the vacuum has a non-trivial structure of the chiral condensate $\bar{q}q$,
which is a similar structure to the Higgs boson known for the electroweak symmetry
breaking~\cite{PhysRevLett.13.508} in the Standard model.

The expectation value of the chiral condensate $|\qq|$ is an order parameter
of the chiral symmetry, 
which is expected to decrease at high temperatures and/or high matter densities
for partial restoration of the chiral symmetry.
So far, $\qq$ has been well investigated at high temperatures.
Numerical calculations in the framework of lattice quantum chromodynamics (QCD) have been yielding stimulating results~\cite{DeTar09,Fu20}.
Experimentally, extremely high-temperature conditions have been explored
by head-on collisions of nuclei at ultra-relativistic energies
to generate quark-gluon-plasma~\cite{RHIC}.
In contrast, experimental knowledge at finite densities
is limited. Lattice QCD calculations encounter ``sign problems'',
general computational difficulties in numerical calculations of Fermion systems, 
in this region~\cite{Fukushima10}.
Experimental measurements have been conducted either by
collisions of nuclei at relatively lower energies to form high-density baryonic matter~\cite{Friman11}
or by spectroscopy of hadrons in nuclear matter~\cite{E325,Kim20}.

In terms of the QCD vacuum, the nuclear matter serves as an impurity or a chemical potential
loaded to the vacuum. Spectroscopic measurements of meson-nucleus bound systems offer
opportunities to research the ``medium effect'' on the fundamental symmetries of the QCD vacuum~\cite{Jido08}.
Although the chiral condensate is invisible,
a spectrum of the masses, the self-energies in the scalar interactions,
reflects the underlying structure of the vacuum. For instance,
the mass differences between chiral-partner hadrons
such as $\rho$ and $a_1$ mesons or nucleon and $N^\ast(1535)$ baryon resonance
become smaller as the chiral symmetry restores.
Likewise, the mass spectrum of the lightest pseudoscalar nonet,
i.e. $\pi$, K, $\eta$, and $\eta'$,
is expected to be changed significantly in the chiral symmetry restoration phase 
since the presently observed masses in the vacuum strongly reflect the broken 
chiral symmetry and the axial U(1) anomaly~\cite{Klimt90,Nagahiro13}.

When the meson and the nuclear wavefunctions overlap largely,
we obtain information on their $s$-wave interaction, which
is modified by the wavefunction renormalization in the medium effect.
We can investigate the density dependence of the interaction
and derive information of the partial restoration of the chiral symmetry in the nuclear matter~\cite{Jido08,Friedman19}.
For instance, Kaonic atoms and nuclei may provide information of
the $\bar{s}s$ component of the chiral condensate.
Spectroscopy experiments are in preparation to measure the bound states of various mesons
in nuclei~\cite{Ajimura19,Yamaga20,Ishikawa17,Tanaka16}.

Pionic atoms, bound systems of a $\pim$ and a nucleus, provide
quantitative information of $\qq \equiv \Braket{\bar{u}u+\bar{d}d}$ in nuclei using the
$\pim$ as a probe~\cite{Suzuki04}. A major part of the pion wavefunction
is located near the surface of the nucleus in a counter balance between the attractive Coulomb interaction
and the repulsive $s$-wave pion-nucleus strong interaction~\cite{Gilg00,Itahashi00}.

The pion-nucleus interaction is phenomenologically described by an optical potential
of the Ericson-Ericson formulation~\cite{Ericson66} presented in Section METHODS SUMMARY,
which reproduces many of pionic atom data~\cite{Batty97,Friedman07}.
For pionic atoms with relatively heavy nuclei,
the binding energies and the widths of the pionic $1s$ or $2p$ states
are predominantly determined by the $s$-wave
interaction whereas the higher orbitals are mostly determined
by the $p$-wave part~\cite{Yamazaki96,Itahashi00,Geissel02,Yamazaki12}.
The $p$-wave part was studied by making a fit to the existing
pionic X-ray spectroscopy data measured in the transitions between the higher orbitals.
In the $s$-wave part, the isovector potential is the leading-order part of the potential,
which is proportional to the density
difference of neutrons $\rho_n(r)$ and protons $\rho_p(r)$ while
the density-proportional isoscalar potential vanishes~\cite{Tomozawa66,Weinberg66,Batty97}.

Due to the medium effect,
the $s$-wave interaction is modified in the nuclear matter~\cite{Kolomeitsev03}. 
This effect on the leading-part isovector interaction is
expressed as density dependence of the isovector parameter $b_1(\rho)$,
which has been intensively discussed recently based on the $\pi$-nucleus scattering data~\cite{Friedman05} and
the pionic atom data~\cite{Friedman19}.

A pioneering experiment of pionic atoms
investigated the in-medium pion-nucleus interaction and the chiral condensate
in the nuclear matter based on the in-medium Gell-Mann--Oaks--Renner relation~\cite{Suzuki04}.
The ratio of the in-medium chiral condensate to that in vacuum was estimated to be
$\qq(\rho_c)/\qq(0) \sim 67\%$.
This successful $\qq$ estimation has been used as an experimental basis in
discussions of the QCD at low energies.
However, the reported value above has no errors associated
and new data have been awaited for.

Theoretical progress has been recently made in studies of the enhanced isovector interaction
and the $\qq$ via the in-medium Glashow-Weinberg relation~\cite{Jido08}.
The isovector parameters $b_1(\rho)$ in the medium and $b_1(0)$  in the vacuum are
model-independently related
to the ratio of the chiral condensate in the vacuum $\qq(0)$ and the medium $\qq(\rho)$ by
\begin{equation}
\frac{\qq(\rho)}{\qq(0)} \simeq \left(\frac{b_1(0)}{b_1(\rho)}\right)^{1/2}\left(1-\gamma\frac{\rho}{\rho_c}\right),
\label{eqn:qqb1}
\end{equation}
where $\rho_c \equiv 0.17$ fm$^{-3}$ is the normal nuclear density and the coefficient $\gamma = 0.184 \pm 0.003$.

\section*{Spectroscopy of pionic atoms}
\begin{figure}[hbtp]
\centering{
  \includegraphics[width=8.6cm]{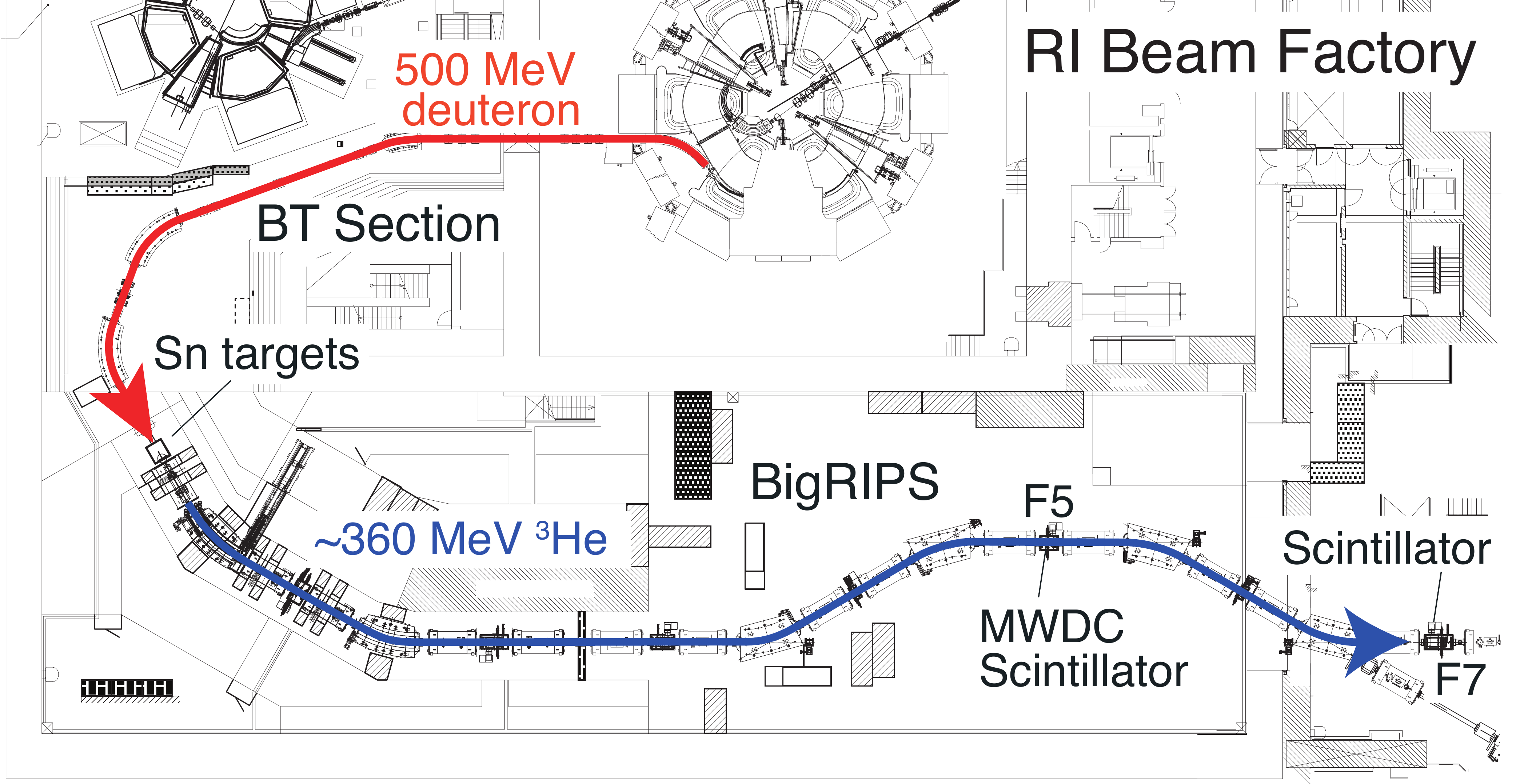}
  \caption{\textbf{Experimental layout.} 500 MeV deuteron beam shown in the red arrow was accelerated by an accelerator complex at the RI Beam Factory and
    impinged on Sn targets. The $\helium$ particles from the  $\reaction$ reactions shown in the blue arrow were momentum-analyzed by the BigRIPS spectrometer.
    Sets of Multi-Wire Drift Chambers (MWDC) were installed at the focal plane F5 to measure the $\helium$ tracks. Two sets of plastic scintillation counters were installed at
    the focal planes F5 and F7.
  }
  \label{Fig:BigRIPS}
}
\end{figure}

Here we report the determination of $\qq(\rho)/\qq(0)$ with error bars
based on the data of pionic Sn atoms and recent theoretical progress.
The experimental setup is schematically depicted in Fig.~\ref{Fig:BigRIPS}.
The experiment was conducted in 2014
at the RI Beam Factory, RIKEN~\cite{Kubo03} using the high intensity deuteron $d$ beam of $\sim 10^{12}$/s
with an energy of $501.3 \pm 0.2$ MeV impinging on a 12.5 $\pm\ 0.5$ mg/cm$^2$ thick $\Sn{122}$ target.
We measured the missing mass of the $\Sn{122}\reaction$ nuclear reactions and observed the excitation spectra near
the $\pim$ production threshold. The produced pionic atoms are coupled with neutron hole states  in the ${}^{121}{\rm Sn}$ nucleus.
Relevant neutron hole states with the excitation energy $E_n(n'j'l')$
are $2d_{3/2}$ (0.0 MeV), 
$3s_{1/2}$ (0.06034 MeV), 
$2d_{5/2}$ (1.1212 MeV), and
$2d_{5/2}$ (1.4035 MeV)~\cite{Ohya10}.

We established a method of formation and observation 
of pionic atoms to achieve high spectral resolution
in the nuclear reactions with the scattering angles $\theta$ up to several degrees~\cite{Nishi18}.
The major contribution to the spectral resolution
is the momentum distribution of the deuteron beam~\cite{Nishi13,Nishi18}.
The momentum width of the incident deuteron beam was as large as 0.03\% ($\sigma$).
We applied the ion-optical ``dispersion-matching'' technique with the diagnostics
for instantaneous feedback and optimization.
The beam line was tuned dispersively on target and achromatically in the focal plane. This avoids the large momentum
spread of the beam to affect the spectral resolution~\cite{Fujita02}.

\begin{figure}[hbtp]
\centering{
  \includegraphics[width=8.6cm]{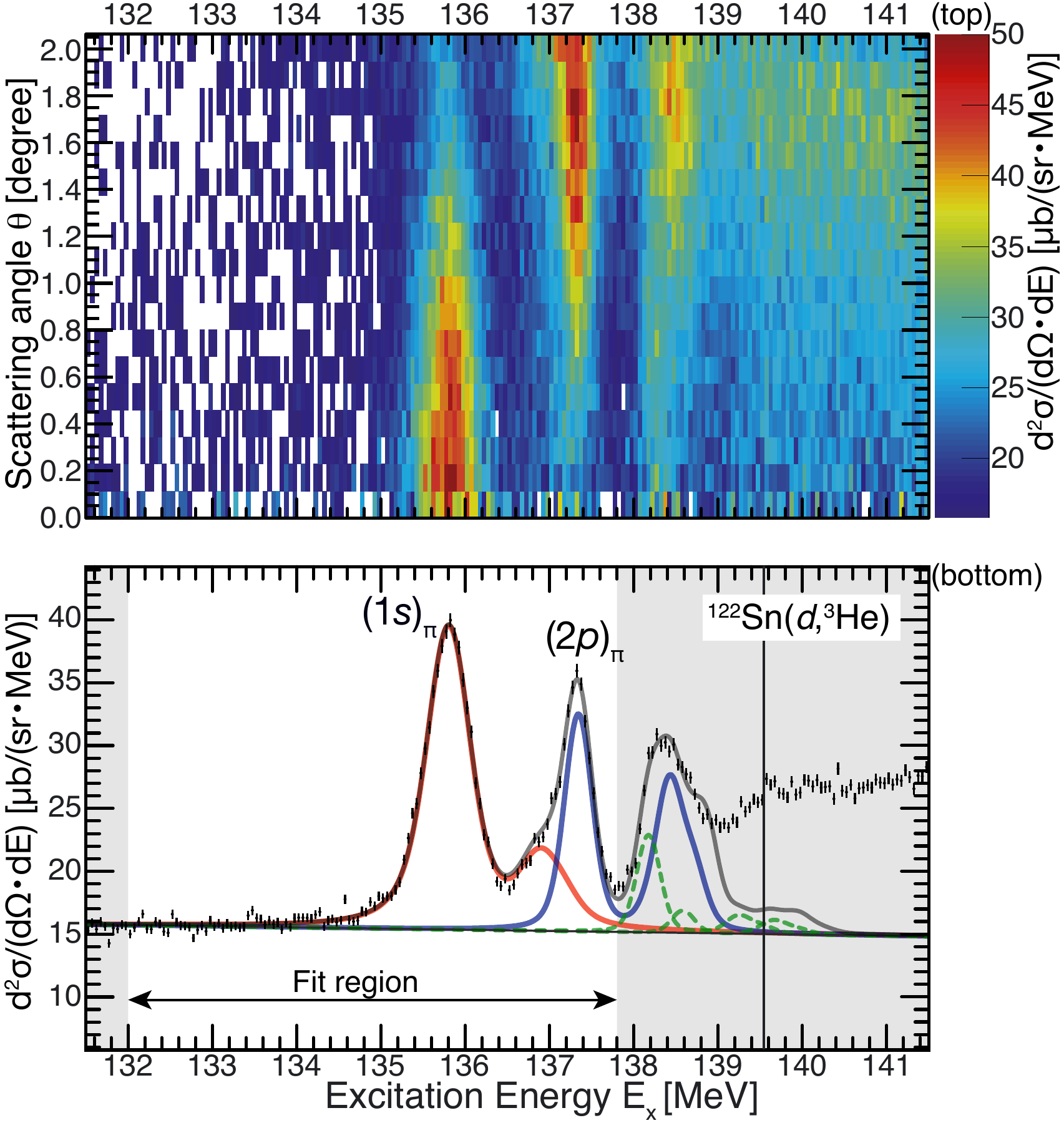}
  \caption{\textbf{Measured spectra.}
    (top) Scattering angle $\theta$ dependence of the double differential cross section.
    (bottom) Measured excitation spectrum of the $\reaction$ reaction for the scattering angle $<1.5$ degrees.
    The tiny vertical bars show the statistical errors $(\sigma)$.
    The most prominent peak near $E_x \sim 135.7$ MeV is assigned to formation of pionic $\Sn{121}$ atoms in the $(1s)_\pi$ state and a smaller peak
    near $E_x \sim 137.3$ MeV to the $(2p)_\pi$ state. The pion emission threshold is shown by the vertical black line.
    As shown by the gray curve, we have fitted the spectrum in the $E_x$
    region indicated by the arrows and the grey masks
    using $B_\pi$, $\Gamma_\pi$, and the cross sections
    of the $(1s)_\pi$ and $(2p)_\pi$ states
    and a linear background as the free parameters.
    The contributions from the pionic $1s$, $2p$, and the other states are decomposed and shown by the red, blue, and green(dashed) curves, respectively.
  }
  \label{Fig:Spectra}
}
\end{figure}

Figure~\ref{Fig:Spectra} (bottom) shows the measured excitation spectrum of the pionic $\Sn{121}$ atoms
for $\theta< 1.5$ degrees. The tiny vertical bars show the statistical errors. The statistical precision is
much better than in preceding experiments for the whole spectral range~\cite{Suzuki04,Nishi18}, which
is mandatory for detailed studies of the pion-nucleus interaction.
The abscissa is the excitation energy $E_x$ of the reaction products near the $\pim$ emission
threshold represented by the vertical line.
The ordinate is the double differential cross sections of the $\reaction$ reaction.
The energy resolution has parabolic $E_x$ dependence.
The best resolution of 287~keV (FWHM) is obtained near $E_x \sim 138.5$ MeV.

The formation of the $1s$ and $2p$ pionic atoms are observed as distinct peaks.
Comparing the spectrum with the theoretical calculations in Fig.~1 and 4 of Ref.~\cite{Ikeno15},
we find fairly good agreement of the overall shape of the spectra although the
absolute $1s$ strength is smaller as discussed in Ref.~\cite{Nishi18}.
Among the configurations of the pion wavefunctions $(nl)_\pi$
and the neutron holes $(nl)_n^{-1}$,
the largest strengths are from the $(1s)_\pi(3s_{1/2})_n^{-1}$ and
$(2p)_\pi(3s_{1/2})_n^{-1}$states.

Figure~\ref{Fig:Spectra} (top) depicts the $\theta$ dependence of the
pionic atom formation cross section. The ordinate is $\theta$ and the
abscissa is $E_x$. The $1s$ formation cross section peaks at $\theta=0$ and
decreases for larger $\theta$. In contrast, the $2p$ cross section increases
for larger $\theta$.

The experimental spectrum in Fig.~\ref{Fig:Spectra}(bottom)
has been fitted for the region indicated by a linear background and theoretical spectra
in the same manner as in Ref.~\cite{Nishi18}.
The fitting parameters are 
the $1s$ and $2p$ binding energies ($B_\pi$), widths ($\Gamma_\pi$), and formation cross sections and the linear background.
Each contribution of a pionic state and a neutron-hole state
is given by a Voigt function with the Lorentzian width of the pionic
level and the Gaussian width of the experimental resolution.
Theoretical results of the effective number approach in Ref.~\cite{Ikeno15} are used for
the relative strengths of the neutron-hole contributions after incorporating
the recently measured values with the errors in the spectroscopic factors of the neutron hole states~\cite{PhysRevC.104.054308}.
The gray curve is the fitting result with the $(1s)_\pi$ spectrum shown as red curve,
$(2p)_\pi$ as blue curve, higher orbitals as dashed green curve, and a linear background.
The fitting $\chi^2$/n.d.f. is 231.3/108.

Table~\ref{Table:BG} summarizes the deduced $B_\pi$ and $\Gamma_\pi$ with statistical and systematic errors.
We have also evaluated the differences for the $1s$ and $2p$ values since
a part of the systematic errors is common to the $1s$ and $2p$ values.
Taking the differences we achieve much better accuracy for $B_\pi(1s)-B_\pi(2p)$.
We briefly discuss the evaluated systematic errors in Section METHODS SUMMARY.

\begin{table}[hbtp]
  \begin{tabular}{cccc}
\hline
\hline
& [keV] & Statistical & Systematic \\
\hline
$B_\pi(1s)$ & 3830 & $\pm 3$ & $+78-76$\\
$B_\pi(2p)$ & 2265 & $\pm 3$ &$+84-83$\\
$B_\pi(1s)-B_\pi(2p)$ & 1565 & $\pm 4$ & $\pm 11$\\
\hline
$\Gamma_\pi(1s)$ & 314 & $\pm 11$ & $+43-40$\\
$\Gamma_\pi(2p)$ & 120 & $\pm 12$ & $+49-28$\\
$\Gamma_\pi(1s)-\Gamma_\pi(2p)$ & 194 & $\pm 16$ & $+31-42$\\
\hline
\hline
  \end{tabular}
  \caption{
    \textbf{Deduced \boldmath$B_\pi$ and $\Gamma_\pi$ of pionic $\Sn{121}$ atom.}
    The binding energies $B_\pi$ and widths $\Gamma_\pi$
    of the $1s$ and $2p$ states are tabulated.
    Differences of the $1s$ and $2p$ values are also shown.
    ``Statistical'' denotes the statistical errors $(\sigma)$ and ``Systematic''
    the systematic errors.
  }
  \label{Table:BG}
\end{table}

\section*{Pion-nucleus interaction}
Now we discuss the deduction of the $\pim$--nucleus potential parameters.
We have calculated binding energies and widths for a certain set of
potential parameters and compared them with our observables $B_\pi$ and $\Gamma_\pi$.
We have evaluated the likelihood defined as the differences between the
calculated values and the observables
to determine the best-fit parameters in a statistical way.
The high-quality information of the $1s$ and $2p$ bound states in Sn nuclei are
simultaneously used.
In order to focus on the isovector parameters
and set constraints on the isoscalar parameters $b_0$ and Re$B_0$, we have introduced the data of the
light spherical nuclei of ${}^{16}{\rm O}$, ${}^{20}{\rm Ne}$, and ${}^{28}{\rm Si}$~\cite{Batty97},
for which the nuclear distributions fulfill the condition $\rho_p(r) = \rho_n(r)$. Note that the isovector
parameter is hardly affected by the data of these spherical nuclei.

In the analysis, we have carefully examined the Ericson-Ericson formulation
of the optical potential~\cite{Ericson66} and made an update based on considerations of the $\pim$ absorption
processes in the nuclear medium. We have employed
$(4/3)(\rho_p^2(r) + 2\rho_n(r) \rho_p(r))$ instead of the conventional $\rho^2(r)$ as the absorption term
to take into account the dominance of the $\pim$ absorption by two protons or by a pair of a neutron and a proton.
We have applied the Lorentz-Lorenz parameter $\xi=1$
for the short-range correlations of nucleons following the analysis of Ref.~\cite{Friedman03} and made a comparison
with the case using $\xi=0$.
We have fixed the $p$-wave parameters to the ``Global 2'' parameters shown in Table 2 of Ref.~\cite{Friedman03}.

For the evaluation of $\rho_n(r)$ of $\Sn{121}$
we have employed recent systematic data of the proton elastic scattering reaction.
In preceding studies, large ambiguities were observed in $\rho_n(r)$ as discussed in Refs.~\cite{Suzuki04,Friedman07}.
There, simple two-parameter Fermi models~\cite{Fricke95} were applied assuming skin-type and halo-type distributions.
In our work, we have adopted the high precision data measured
at RCNP, Osaka University in Ref.~\cite{Terashima08},
interpolated between $\Sn{120}$ and $\Sn{122}$ nuclei to obtain $\rho_n(r)$ for
the $\Sn{121}$ nucleus.
Using this procedure, we have achieved remarkably small uncertainties in $\rho_n(r)$.
For $\rho_p(r)$, we have used the data from the electron scattering experiments~\cite{Ficenec72}.
We have calculated the overlap between the nuclear density
and the pionic wavefunctions. 
The overlap surges near the nuclear surface
with a maximum at the effective density $\rho_e = 0.58 \rho_c = 0.098$ fm$^{-3}$.
The interaction parameters being determined by the fit
represents the values at $\rho_e$.

We have taken into account residual interactions between the pion and the nucleus with a neutron hole
to evaluate the effect of the diagonalization of the whole Hamiltonian of the isolated quantum object,
the pion-nucleus system. This effect had been neglected in previous studies.
The residual interactions have been introduced in the present analysis as corrections
evaluated in the same way as in Ref.~\cite{Nose-Togawa05}.
The numerical evaluation of the formation cross sections by the Green's function method~\cite{Ikeno15}
and the effective number method~\cite{Hirenzaki91} are compared and the differences are also taken into account.

Table~\ref{Table:b1} summarizes the results of the above state-of-the-art analyses.
These methods and improvements
are directly compared with the case of  ``classical'' approaches
in earlier publications~\cite{Suzuki04}. The largest differences are found
in the adoption of the measured neutron density distributions denoted as Osaka.
In total, we have $b_1$ shifted substantially by $0.0211\ m_\pi^{-1}$, where $m_\pi = 139.57$ MeV$/c^2$ is the pion mass.
This shift must be kept in mind for comparison with the earlier publications.

We have thus deduced the optical potential parameters by a likelihood fitting of the values of
$B_\pi(1s)-B_\pi(2p), B_\pi(1s), \Gamma_\pi(1s)$ and $\Gamma_\pi(2p)$ as observables.
For the cancellation of the systematic errors, this combination of the observables has large significance.
We have calculated the likelihood of $B_\pi$ and $\Gamma_\pi$ as a function of $b_1$ and Im$B_0$
taking into account the statistical and systematic errors and their correlations,
which largely improved the precision of the deduction.
The deduced values are $b_1 = (-0.1163 \pm 0.0056)\  m_\pi^{-1}$ and  Im$B_0 = (0.0473 \pm 0.0013)\  m_\pi^{-4}$.
The best-fit $b_0$ and Re$B_0$ are $-0.0225 \ m_\pi^{-1}$ and $-0.0220\ m_\pi^{-4}$, respectively.
The fitting $\chi^2$/ndf is 1.7/6.

\begin{table}[hbtp]
  \begin{tabular}{ccccc|cc}
\hline          
\hline
$\xi$ & $\rho_n(r)$ & Abs. & C.S. & Res. & $b_1 [m_\pi^{-1}]$ & Im$B_0 [m_\pi^{-4}]$ \\
\hline

0 & 2pF & $\rho^2$ & Neff & $-$ & $-0.0952$ & $0.0469$ \\
1 & 2pF & $\rho^2$ & Neff & $-$ &$-0.0945$ & $0.0472$ \\
1 & Osaka & $\rho^2$  & Neff & $-$ & $-0.1096$ & $0.0472$ \\
1 & Osaka & $pp+2np$ & Neff & $-$ & $-0.1116$ & $0.0473$ \\  
1 & Osaka & $pp+2np$ & Green & $-$ & $-0.1148$& $0.0473$ \\ 
1 & Osaka & $pp+2np$ & Green & \checkmark & $-0.1163$ & $0.0473$ \\ 

\hline
  \end{tabular}
  \caption{
    \textbf{Magnitude of effects on \boldmath$b_1$ and Im$B_0$ parameters by the methods.}
    The best fit values of $b_1$ and Im$B_0$
    are displayed in the right columns with applying the methods
    listed in the left columns.
    $\xi$ denotes the Lorentz-Lorenz factor of the
    short range nucleon correlation~\cite{Batty97}, 2pF the neutron density of two-parameter-Fermi model
    in Ref.~\cite{Fricke95}, Osaka the neutron density distribution determined in Ref.~\cite{Terashima08},  Abs. the imaginary-term
    formulation
    in the Ericson-Ericson potential either with $\rho^2$ for the Im$B_0 \rho^2(r)$ form
    or  with $pp+2np$ for the $(4/3)$Im$B_0(\rho^2_p(r)+2\rho_n(r)\rho_p(r))$ form,
    C.S. the adopted numerical method to analyze the
    formation cross section by the Green's function method (Green)~\cite{Ikeno15} or by
    an effective number approach (Neff)~\cite{Hirenzaki91},
    and Res. the residual interaction of the neutron hole state~\cite{Nose-Togawa05}.
  }
\label{Table:b1}
\end{table}

\section*{In-medium Chiral Condensate}
We have deduced the $\pim$-nucleus isovector parameter
$b_1 = (-0.1163 \pm 0.0056)\  m_\pi^{-1}$ with unprecedented precision and accuracy.
In order to compare the deduced $b_1$ value with the preceding result in Ref.~\cite{Suzuki04}
in a consistent way, we have applied the revised methods tabulated in Table~\ref{Table:b1}
to the preceding result and obtained $b_1 =( -0.136 \pm 0.007)\ m_\pi^{-1}$,
which exhibits a discrepancy of about $2\sigma$'s from the present result.
The above deduced $b_1$ is enhanced by $34 \pm 7\%$
compared to the isovector strength in the vacuum of
$b_1 = (-0.0866 \pm 0.0010)\  m_\pi^{-1}$ deduced 
from the high precision X-ray spectroscopy of the pionic hydrogen and deuterium~\cite{Hirtl:2021tb}.
This enhancement marginally agrees with the calculated values
of $\sim 30\%$ in the chiral perturbation
theory~\cite{Chanfray03}.

Now, we discuss the deduction of $\qq$ in the nuclear medium by analysis of the pion-nucleus interaction and its enhancement due
to the wavefunction renormalization.
The enhancement of $|b_1|$ 
is due to the partial restoration of the chiral symmetry in the nuclear medium.
The value of $b_1 = (-0.1163 \pm 0.0056)\  m_\pi^{-1}$ that we have achieved
is directly translated to $\qq$ at the effective density
$\rho_e$  by Eqn.~(\ref{eqn:qqb1}),
so that we obtain 
$\qq(\rho_e)/\qq(0) = 77 \pm 2 \%$.

\begin{figure}[hbtp]
\centering{
   \includegraphics[width=8.6cm]{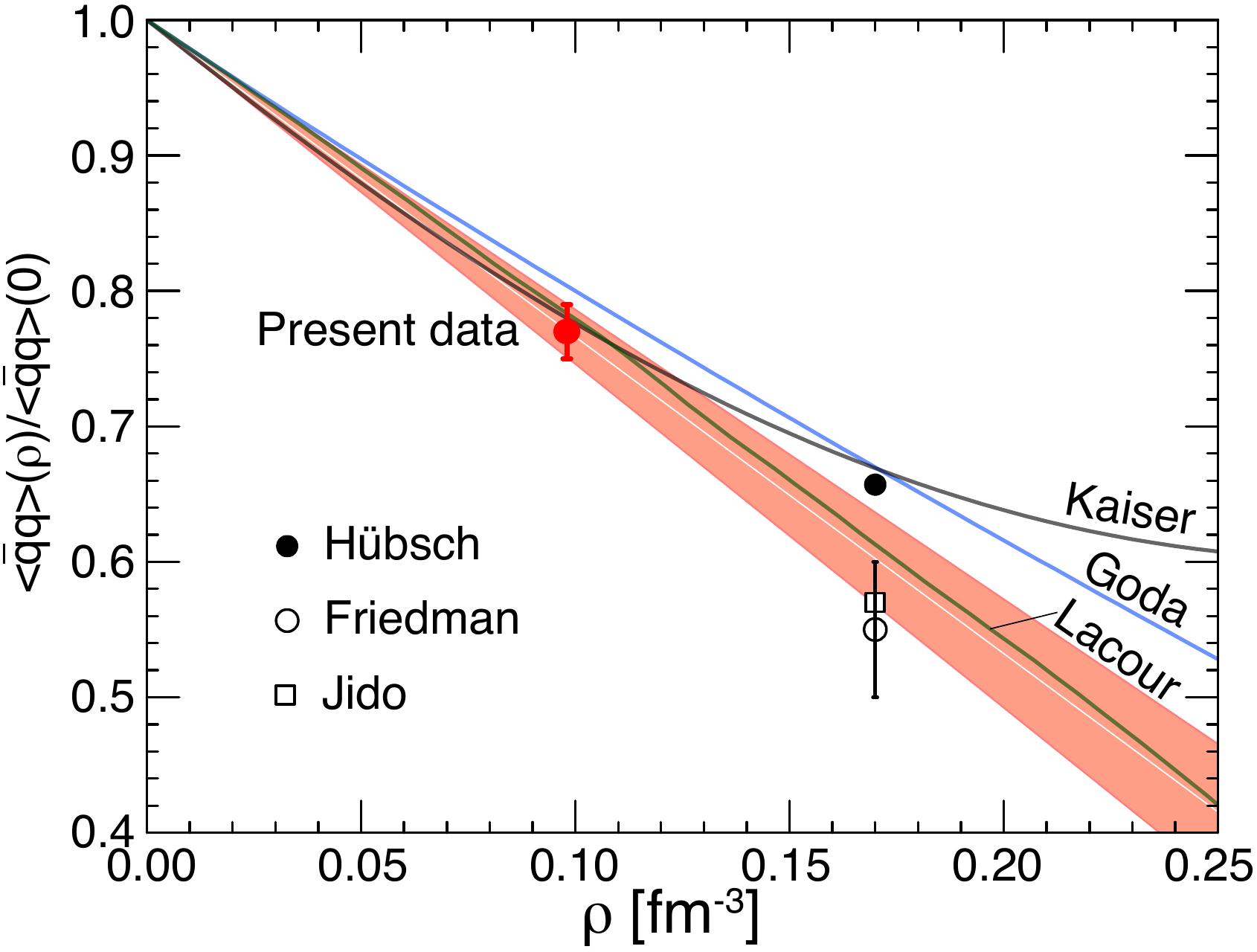}
  \caption{
    \textbf{The deduced in-medium \boldmath$\qq(\rho)$ normalized to $\qq(0)$ in vacuum.}
    The abscissa is the density and the ordinate the ratio of $\qq(\rho)$ in
    the medium to $\qq(0)$ in vacuum.
    The result of the present experiment is shown by the filled red circle with
    the error bars $(\sigma)$.
    The red coloured region with the white line shows the linear extrapolation of the present data with the errors.
    Theoretical results are presented for H\"ubsch~\cite{Huebsch21},
    Friedman~\cite{Friedman19} with the error bars $(\sigma)$, Jido~\cite{Jido08}, Kaiser~\cite{Kaiser08},
    Goda~\cite{Goda13}, and Lacour~\cite{Lacour_2010}.
  }
  \label{Fig:chisq}
}
\end{figure}

Figure~\ref{Fig:chisq} shows the presently deduced $\qq(\rho_e)/\qq(0)$ of $77 \pm 2 \%$
by the filled red circle with the error bars.
The ordinate is the ratio of $\qq(\rho)$ in the medium to $\qq(0)$ in the vacuum and the abscissa is the density $\rho$.
The red coloured region shows the extrapolation of the present data
by assuming the linear density dependence of $\qq(\rho)$.
We obtain $\qq(\rho_c)/\qq(0) = 60 \pm 3\%$ at the normal nuclear density.
For comparison, theoretical results of Refs.~\cite{Jido08}, \cite{Friedman19},
and \cite{Huebsch21} are presented with labels of Jido, Friedman, and H\"ubsch, respectively.
The present $\qq$ ratio of $60 \pm 3\%$ at $\rho_c$ agrees
with calculations of 57\% in Ref.~\cite{Jido08} and $55\pm5\%$ in Ref.~\cite{Friedman19} and slightly deviates
from the recent estimate of 65.7\% in Ref.~\cite{Huebsch21}. 
A gray curve denoted ``Kaiser'' is a theoretical calculation taken from Fig.~5 of Ref.~\cite{Kaiser08}
for the case of $m_\pi=135$ MeV, the blue curve ``Goda'' from Fig.~7 of Ref.~\cite{Goda13},
and the green curve ``Lacour'' from Fig.~8(right) of Ref.~\cite{Lacour_2010}.
The present value agrees with the theoretical curves of Refs.~\cite{Kaiser08} and ~\cite{Lacour_2010}
and slightly deviates from the curve of Ref.~\cite{Goda13}.

For advanced studies, we are preparing systematic measurements
of pionic atoms to deduce the density dependence of the chiral condensate.
At present, we assume a linear density dependence
as depicted in Fig.~\ref{Fig:chisq}.
It is known that the nuclear densities probed by the pionic atoms depend on
the pionic levels and the nuclei~\cite{Ikeno11}. High-precision systematic
spectroscopy will shed light on the low-energy
high-matter-density sector of the QCD.

\section*{Acknowledgement}
The authors thank the staff of the RI Beam Factory for 
stable operation of the facility.
This experiment was performed at RI Beam Factory
operated by RIKEN Nishina Center and CNS, University of
Tokyo. This work is partly supported by MEXT Grant-in-Aid
for Scientific Research on Innovative Areas
(No. JP22105517, No. JP24105712, and No. JP15H00844),
JSPS Grant-in-Aid for Scientific Research (B) (
No. JP16340083 and No. JP18H01242) (K.I.), (A) (No. JP16H02197) (K.I. and T.U.), and
(C) (No. JP24540274 and JP16K05355) (S.H.),
Grant-in-Aid for Early-Career Scientists (No. JP19K14709) (N. Ikeno),
Grant-in-Aid for JSPS Research Fellow (No. JP12J08538) (T.N.),
JSPS Fund for the Promotion of Joint International Research
(Fostering Joint International Research (B)) (No. JP20KK0070) (K.I.),
Institute for Basic Science (IBS-R031-D1) (D.A.), 
the Bundesministerium f\"ur Bildung und Forschung (H.G., E.H., H.W.), and the
National Science Foundation through Grant No. Phys-0758100,
and the Joint Institute for Nuclear Astrophysics
through Grants No. Phys-0822648 and No. PHY-1430152
(JINA Center for the Evolution of the Elements) (G.P.A.B.).

\section*{Author Contributions Statement}
T.N. and K.I. designed experimental concepts
and performed experiments, developed detectors and ion optics, analyzed data, performed
theoretical calculations, and wrote the paper;
G.P.A.B., M.D., H.F., N.Fukuda, N.Fukunishi, H.G., E.H., K.Kusaka, N.S., H.S., K.S., H.T., Y.K.T., T.U., Y.W., H.W.
developed ion optics and performed experiments;
S.H., N.Ikeno. designed experimental concepts, performed theoretical calculations and wrote the paper;
N.N-T. performed theoretical calculations and wrote the paper;
H.M. developed detectors; H.N. and M.I. designed experimental concepts;
All the others performed experiments.

\section*{Competing Interests Statement}
The authors declare no competing financial interests.

\newpage

\section*{Methods Summary}
We measured the excitation energy of the $\Sn{122}\reaction$ reactions
near the $\pim$ production threshold.  A neutron is picked up by the
incident deuteron and a $\helium$ is emitted while a $\pim$ is
transferred to the target Sn nucleus. The $\pim$ has a small momentum
of $q \leq 20$ MeV$/c$. This enhances the capture cross section by the
target nuclei to form a pionic atom.

The outgoing $\helium$ in the reaction was mainly identified by the time-of-flight measured
between F5 and F7 focal planes and the energy-loss
measured by the scintillation counters. Typical counting rates of $\helium$ and background
protons were about 100 Hz and $\sim 1$ MHz, respectively. We achieved nearly
perfect identification of the $\helium$. The contamination from other particles is negligible.
We used the fragment separator BigRIPS as a spectrometer.
We have made detailed analysis of the ion-optical properties of the BigRIPS spectrometer
and made corrections of the higher-order aberrations up to the third order.
The uncertainties in the corrections were taken into account in the evaluation of the systematic errors.

The absolute scale of the excitation energy was calibrated by using the
two-body $\pi^0$ production reaction on hydrogen
H$\reaction\pi^0$. The reaction produces nearly mono-energetic
$\helium$ at forward 0 degree angles.  We developed detailed
simulations, reproduced the angular dependence of the $\helium$
distributions, and obtained a good calibration of the excitation
energy. The systematic error associated with the absolute
excitation energy has been evaluated to be $\sim$ 5 keV.

The measured $\helium$ momenta have been correlated to the excitation
energies.  We improved the experimental resolution of the measured
momentum by employing a specially developed ion-optical setting to
eliminate the contribution from the momentum spread of the incident beam.
In this ``dispersion matching'' setting,
the beam is momentum dispersed on target so that the beam is achromatic
at the focal plane of the BigRIPS spectrometer section (see Fig.~\ref{Fig:BigRIPS}).
We have realized the dispersion matching setting by developing a diagnostics method
for instantaneous feedback.
Thus, the contribution of the momentum spread is mostly eliminated
in the spectrum.  The remaining contributions to the spectral resolution are given by
the ion-optical aberrations of the spectrometer and the multiple scattering
in the target and the detectors. The FWHM resolution is well fitted
in the range of $E_x=$ [133,138] MeV by a third-order polynomial
function of $E_x$ in MeV unit: $(286.6 + 22.58 (E_x - E_x^0)^2 + 1.228
(E_x - E_x^0)^3)$ keV, where $E_x^0 = 138.5$ MeV.

Theoretically, the formation
cross section of the pionic atoms is described well by an effective
number method or by the Green's function method. While
the Green's function method can accurately describe the bound states
with non-zero decay widths,
its computational cost is high.
In contrast, the effective number method requires relatively low computational costs.
We have compared the results
using both methods with the same parameters and evaluated
the difference presented in Table~\ref{Table:b1}.
The excitation spectrum consists of contributions from 
pionic and neutron hole states. Each contribution was calculated
based on the spectroscopic factors of the neutron states, the integrated overlaps of the pionic and neutron wavefunctions,
the pionic binding energies and widths, and the neutron
separation energies.  We assumed 20\% errors in the spectroscopic factors.

The Ericson-Ericson formulation of the pion--nucleus optical potential
$U_{\rm opt}(r)$ is described as
\begin{eqnarray*}
  2\mu U_{\rm opt}(r) &=& U_s(r) + U_p(r) \label{eqn:E-E opt}\\
  U_s(r) &=& -4\pi [\epsilon_1\{b_0\rho(r)+b_1 \Delta\rho(r) \} + \epsilon_2  B_0\rho^2(r)]\\
  \rho(r) &=& \rho_n(r)+\rho_p(r)\\
  \Delta \rho(r) &=& \rho_n(r)-\rho_p(r)
\end{eqnarray*}
where $\epsilon_1 = 1 + \mu/M = 1.147, \epsilon_2 = 1 + \mu/2M =1.073$,
with $\mu$ being the pion-nucleus reduced mass and 
$M$ being the nucleon mass~\cite{Ericson66,Batty97}.  $U_s(r)$ and $U_p(r)$ 
denote the $s$-wave and $p$-wave parts, respectively.
$b_0$, $b_1$ and $B_0$ are the $s$-wave
isoscalar, isovector and a complex parameters, respectively.
A wide range of the pionic atom data is known to be fitted well by a
set of parameters.
Particularly, the $p$-wave part is relatively well determined from the pionic atoms
in the outer orbitals. Note that binding energies and widths of the inner $1s$ or $2p$ orbitals are scarcely
affected by the $p$-wave parameters.
While the isoscalar part $b_0$ is
known to be small, the isovector interaction $b_1$ is the
leading-order term.  The imaginary term of $B_0$ describes
the $\pim$ absorption in the nuclei.  In the present analysis, taking
into consideration that the $\pim$ is absorbed either by proton-proton
or by proton-neutron pairs and not by neutron-neutron pairs, we have
replaced Im$B_0\rho^2(r)$ by Im$B_04/3
(\rho_p(r)^2+2\rho_p(r)\rho_n(r))$. Note that this modification
preserves calculation results for the $\rho_p(r) = \rho_n(r)$ nuclei.

For comparison of the deduced $|\qq|$ with the theoretical values, we have performed
the following calculations. The ratio $\qq(\rho_c)/\qq(0)$ of 55 $\pm$ 5\%
was calculated for the $\pi N\sigma$ term $\sigma_{\pi N} = 57\pm7$ MeV
and the pion weak decay constant $f_\pi = 92.2$ MeV by Eqn. (4) in Ref.~\cite{Friedman19}.
For Refs.~\cite{Kaiser08}, \cite{Goda13}, and \cite{Lacour_2010},
we read the density dependent $|\qq|$ ratio
for $m_\pi = 135$ MeV  in Fig.~5 of Ref.~\cite{Kaiser08},
that for ``Up to NNLO'' in Fig.~7 of Ref.~\cite{Goda13}
and that for the ``symmetric nuclear matter'' in Fig. 8 of Ref.~\cite{Lacour_2010}, respectively.
For Ref.~\cite{Jido08}, 57\% was calculated based on
the pion--nucleus scattering data.
For Ref.~\cite{Huebsch21}, we applied the $|\qq|$ reduction of 34.3\%
at $\rho_c$ neglecting the small difference in the $\rho_n/\rho_p$ ratio.

\section*{Data Availability}
Raw data were generated at the RI Beam Factory. Derived data
supporting the findings of this study are available as Source Data.

\section*{Code Availability}
The computer codes used to generate results are available
from the corresponding author upon reasonable request.
\end{document}